\documentclass[%
 reprint,
 amsmath,amssymb,
 aps,
]{revtex4-2}

\usepackage{graphicx}
\usepackage{dcolumn}
\usepackage{bm}
\usepackage{bm}
\usepackage[utf8]{inputenc}
\usepackage{xcolor}

\begin{document}

\preprint{APS/123-QED}

\title{A Novel Arm-Length Stabilization Scheme for Gravitational-Wave Detectors with AlGaAs/GaAs Coated Mirrors}

\author{Jian Liu}
\email{jian.liu@uwa.edu.au}
\author{Juntao Pan}
\author{Carl Blair}
\author{Chiara Di Fronzo}
\author{Li Ju}
\author{Chunnong Zhao}
\affiliation{%
 OzGrav, University of Western Australia, Crawley, Western Australia 6009, Australia
}%

\author{Sheon S. Y. Chua}
\author{Bram J. J. Slagmolen}
\affiliation{
 OzGrav, Australian National University, Canberra, Australian Capital Territory 0200, Australia
}%

\author{Nutsinee Kijbunchoo}
\author{David Ottaway}
\affiliation{%
 OzGrav, University of Adelaide, Adelaide, South Australia 5005, Australia
}%

\date{\today}

\begin{abstract}
 The arm length stabilisation system is employed in gravitational-wave detectors to reduce the velocity of the mirrors such that the arm cavities can be brought onto resonance in a controlled manner required to attain the detector operating point. For future upgrades of current gravitational wave detectors such as $\rm A^{\#}$, which will incorporate AlGaAs/GaAs coatings, the current frequency-doubled arm length stabilisation system is unsuitable due to excessive absorption of the frequency-doubled 532\,nm beam by the AlGaAs/GaAs coating. We propose a novel multi-wavelength arm length stabilisation scheme that uses both frequency-doubled and frequency-tripled beams. The 1596\,nm auxiliary locking beam is outside the absorption bands of AlGaAs/GaAs coating. It is frequency-tripled to 532\,nm and phase-locked with the 1064\,nm science laser through its second harmonic at 532\,nm. In a tabletop setup, we experimentally demonstrated the stable cavity detuning and robust cavity locking transition by controlling the 1596\,nm laser and 1064\,nm laser phase locked loop. This demonstration confirmed that the proposed novel arm length stabilisation scheme is compatible with future upgrades or third-generation gravitational wave detectors that use AlGaAs/GaAs-coated test masses. 
\end{abstract}

\maketitle


\section{Introduction}
Current operating ground-based interferometric gravitational-wave detectors (GWDs), including Advanced LIGO~\cite{aasi2015advanced}, Advanced Virgo~\cite{acernese2014advanced}, and KAGRA~\cite{aso2013interferometer}, all enhance their sensitivity by using multiple coupled optical cavities. At the detectors' operating point, these cavities are all maintained on resonance (locked) with control systems. Although mirrors from these cavities are isolated from the ground motion via vibration isolation systems, residual large mirror displacement at low frequencies results in the mirrors swinging through arm cavity resonances. Arm cavity resonance changes the sign of the error signals of power recycling and signal recycling cavities, which are locked before the arm cavities are locked.  Therefore, it is essential that the arm cavities are maintained off resonance during the auxiliary cavity lock acquisition. Then the arm cavity needs to be brought onto resonance in a controlled manner such that the sign of the recycling cavity control gains can be switched at the critical moment. As there are multiple length degrees of freedom to be controlled, the sensing signals need to be decoupled to allow for a more deterministic lock acquisition procedure.

The arm-length stabilisation (ALS) system is implemented in current ground-based GWDs~\cite{staley2014achieving,mullavey2011arm,akutsu2020arm,izumi2012multicolor}. In the Advanced LIGO ALS setup, an auxiliary 1064\,nm laser is phase-locked to the primary science laser and frequency-doubled to 532\,nm through second-harmonic generation (SHG). The 532\,nm green light is injected from the end test masses of the arm cavities to pre-stabilise the arm cavity lengths using the Pound–Drever–Hall (PDH) locking technique~\cite{drever1983laser}. By applying a frequency offset on the 532 nm light with an acoustic-optic modulator, the arm cavities can be locked independently of the carrier light. This allows the arm cavities to be locked on green but maintained away from the main 1064 nm light resonance. 

Coating thermal noise limits current detector sensitivity around 100 Hz~\cite{capote2025advanced}. Crystalline AlGaAs/GaAs coatings~\cite{cole2013tenfold} are a promising candidate for reducing coating thermal noise. While producing AlGaAs/GaAs coatings at scales above 30\,cm remains a challenge~\cite{cole2023substrate}, they can potentially reduce the coating thermal noise by a factor of five lower than the current Advanced LIGO $\rm Ta_{2}O_{5}/SiO_2$ coating~\cite{harry2007titania}. The planned $\rm A^{\#}$~\cite{gupta2024characterizing} upgrade to Advanced LIGO detector aims to provide significant sensitivity enhancement compared to Advanced LIGO and test technology for future third-generation observatories, including Cosmic Explorer~\cite{evans2021horizon} and Einstein Telescope~\cite{punturo2010einstein}. $\rm A^{\#}$ plans to incorporate AlGaAs/GaAs coatings to mitigate coating thermal noise.

However, this upgrade would render the existing ALS system ineffective, as the AlGaAs/GaAs coatings would absorb the 532\,nm auxiliary beam due to the fact that the bandgap of GaAs is 1.43\,eV, corresponding to a wavelength of approximately 867\,nm~\cite{blakemore1982semiconducting}. Therefore, a new ALS scheme needs to be developed for cavities that use AlGaAs/GaAs coated mirrors. Tanioka~\textit{et~al.}~\cite{tanioka2024experimental} demonstrated that by changing the SHG to an optical parametric oscillator (OPO) in the current ALS system, the frequency-downconverted 2128\,nm beam can be used in arm cavity pre-stabilisation instead of the frequency-doubled 532\,nm beam. At the same time, they also pointed out that the intrinsic frequency noise of the OPO must be properly managed. 

In this paper, we propose and experimentally demonstrate a new multi-wavelength ALS scheme that uses 1596\,nm laser as the auxiliary locking beam. The auxiliary laser is frequency tripled to 532\,nm and the 1064\,nm science laser is frequency doubled to 532\,nm. The two 532\,nm beams are then phase-locked. The phase-locked loop (PLL)~\cite{wang2023simple,kulur2024characterization} controls the relative phase between the 1596\,nm and 1064\,nm lasers. Ramping the offset in the PLL allows a smooth transition between a 1596\,nm and 1064\,nm lock. Section~\ref{theory} introduces the theory of this new ALS scheme, and section~\ref{experiment} details the experimental setup, followed by the experimental results in section~\ref{results}. Finally, section~\ref{discuss} discusses the performance and possible implementation of the ALS scheme in GWDs.

\section{ALS for $\rm A^{\#}$}
\label{theory}
$\rm A^{\#}$ is a planned baseline upgrade of Advanced LIGO after the 5th observation run. The main Advanced LIGO configuration remains the same with a few key new features of $\rm A^{\#}$, including higher arm cavity power up to 1.5\,MW; a higher level of squeezing with 10 dB frequency-dependent squeezing, larger test masses of about 100\,kg and AlGaAs/GaAs crystalline coating to reduce the dominant coating thermal nois. 

To develop an ALS system for $\rm A^{\#}$ with AlGaAs/GaAs coating using the multi-wavelength scheme, an auxiliary laser frequency must be chosen that can resonate in the arm cavities with AlGaAs/GaAs coated mirrors without excessive optical loss. Then, a phase locking scheme must be established such that the auxiliary laser has a stable phase relation to the 1064\,nm science laser. We propose using a 1596\,nm laser as the auxiliary laser and phase-locked to the 1064\,nm science laser through a combination of SHG and third harmonic generation (THG) processes. The system allows the arm cavities to be controlled with respect to the 1596\,nm auxiliary laser and can transition the controls to the main 1064\,nm science laser, fulfilling the ALS requirements.

\begin{figure}[h]
    \centering
    \includegraphics[width=0.8\linewidth]{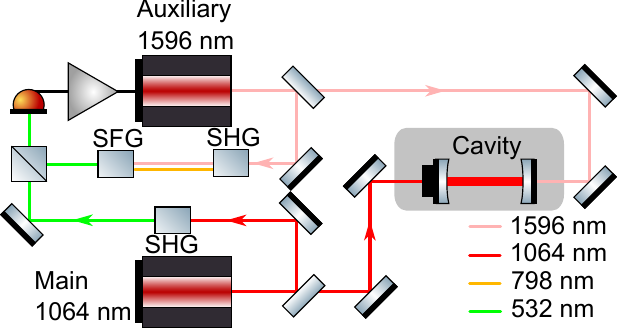}
    \caption{Simplified schematic diagram of the proposed ALS system for $\rm A^{\#}$, showing the frequency doubling of the science laser and frequency tripling of the auxiliary laser, as well as the arm cavity on which the ALS system is applied.}
    \label{fig:AsharpALS}
\end{figure}

The simplified schematic diagram of the proposed ALS system is shown in Fig.\ref{fig:AsharpALS}: a 1596\,nm auxiliary laser is split into two parts through a splitter. Most of the power first goes through a periodically poled potassium titanyl phosphate (PPKTP) SHG crystal, and then through a sum frequency generation (SFG) PPKTP crystal to produce the third harmonic 532\,nm beam. At the same time, the main 1064\,nm science laser is also split into two parts. One part goes to the arm cavity, and the other goes through an SHG to produce the other 532\,nm beam. 

The two 532\,nm beams are then combined together via a beam splitter. The beat signal between the two 532\,nm beams is then then sensed by a high-bandwidth photodetector. A PLL enables the introduction of a precise frequency offset to maintain the beat signal at an arbitrary target frequency. This allows for an arbitrary and stable frequency offset to be established between the 1064\,nm and 1596\,nm beams.

If we denote the frequency of the 532\,nm beam generated from the SHG and SFG as $f_{2\rm nd}$ and $f_{3\rm rd}$, respectively, and ignore the frequency noise associated with the SHG and THG processes~\cite{yeaton2012new}, they are related to frequencies of the 1064\,nm science laser and the 1596\,nm auxiliary laser as
\begin{eqnarray}
f_{\rm 2nd } =&&2\times f_{\rm 1064},\\
f_{\rm 3rd}=&&3\times f_{\rm 1596}.
\end{eqnarray}

When the two 532\,nm beams are phase locked, a deterministic relation between the frequencies of the two beams is established and can be expressed as
\begin{equation}
    f_{2\rm nd} = f_{3\rm rd} + \Delta f,
\end{equation}
with $\Delta f$ being the frequency difference between the two 532\,nm beams, which can be controlled by the PLL frequency offset.

Once the 1596\,nm laser is locked to the cavity, its frequency must be an integer miltiple of the cavity free spectrum range (FSR), $i.e. $
\begin{equation}
    f_{\rm 1596} = \rm N_{1}*{FSR}, N_{1} \in \mathbb{N}.
\end{equation}
Thus, the 1064 nm laser frequency can then be expressed in terms of the cavity FSR as
\begin{equation}
    f_{\rm 1064} = \frac{3\rm N_{1}}{2}*{\rm FSR} + \frac{\Delta f}{2}.
\end{equation}

With the ALS system, the 1064\,nm science laser beam will be on resonance at the same time as the 1596\,nm laser during the locking transition, meaning that the frequency of the 1064\,nm laser will also need to be an integer of the cavity FSR.
\begin{equation}
   f_{\rm 1064} = \rm N_{2}*{FSR}, N_{2} \in \mathbb{N}. 
\end{equation}

This can be achieved by tuning the 532\,nm beam PLL frequency offset $\Delta f$. The value of $\Delta f$ will be determined by the locking conditions of the 1596\,nm laser, with the possible cases being
\begin{equation}
\label{eq:deltaf}
\Delta f=
\begin{cases}
\text{0} &\quad\text{if $\rm N_{1} \bmod 2 = 0$}, \\
\text{$\pm \rm FSR$} &\quad\text{if $\rm N_{1} \bmod 2 = 1$}. \\
\end{cases}
\end{equation}
In these cases, the integer $\rm N_{1}$ and $\rm N_{2}$ are related as
\begin{equation}
\label{eq:interger}
\rm N_{2}=
\begin{cases}
\text{$\rm \frac{3 N_{1}}{2}$} &\quad\text{if $\rm N_{1} \bmod 2 = 0$}, \\
\text{$\rm \frac{3 N_{1} \pm 1}{2}$} &\quad\text{if $\rm N_{1} \bmod 2 = 1$}. \\
\end{cases}
\end{equation}

It can be seen from Eq.(\ref{eq:deltaf},\ref{eq:interger}) that once the auxiliary 1596\,nm laser beam is locked to the cavity, the 1064\,nm laser beams can also be brought to co-resonance in the cavity by tuning the PLL frequency offset $\Delta f$ by no more than the cavity FSR, which for $\rm A^{\#}$, FSR = 37.5\,kHz .

\section{Experimental setup}
\label{experiment}
\begin{figure}[h]
    \centering
    \includegraphics[width=1\linewidth]{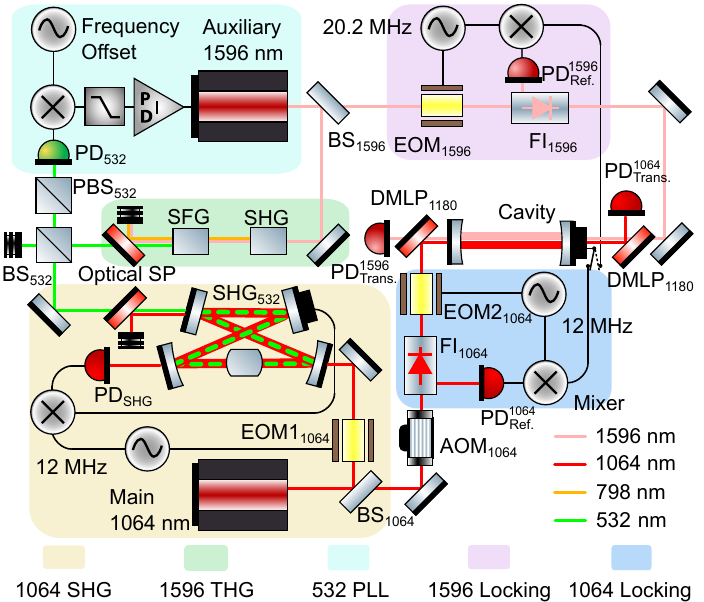}
    \caption{The tabletop experimental setup. Different colored lines represent laser beams of different wavelengths, while black lines indicate electrical connections. The auxiliary laser and science laser locking control signals are both fed back to the cavity PZT mirror, and a switch is used to select the active control path. FI, Faraday isolator; BS, beam splitter; PBS, polarised beam splitter; AOM, acoustic-optical modulator; EOM, electro-optic modulator; PD, photodetector; DMLP, dichroic mirror long-wavelength pass; SP, short-wavelength pass.}
    \label{fig:setup}
\end{figure}

The experimental setup is shown in Fig.\ref{fig:setup}. The whole setup consists of five subsystems, including the 1064\,nm SHG system, the 1596\,nm THG nm system, the 532\,nm phase locking system, the 1596\,nm science cavity locking system and the 1064\,nm science cavity locking system. 

The main science laser is a Mephisto 2000NE single-frequency laser manufactured by Coherent Corp. It outputs a maximum of 2\,W linearly polarised beam at 1064\,nm wavelength with excellent noise performance. Approximately half of the power of the science laser beam is phase modulated at 12\,MHz and coupled into a bowtie-shaped SHG cavity. The SHG cavity is locked via the PDH locking technique~\cite{drever1983laser}, and the PPKTP crystal is housed in a copper oven to maintain a stable temperature at 45\,°C. The SHG cavity can produce a 532\,nm green beam with a power of about 500\,mW. An optical short-wavelength-pass filter is placed after the SHG cavity to block the remaining 1064\,nm beam and allow only the 532\,nm beam to propagate towards a beam splitter, where it beats with the 532\,nm beam generated through the THG process of the 1596\,nm laser. 

The 1596\,nm laser is a custom-made unit provided by Shanghai Precilasers Technology Co., Ltd. It is a fibre-distributed feedback (DFB) laser with a maximum output power of approximately 1.2\,W. Its wide wavelength tuning range of about 1\,nm allows the frequency-tripled beam to spectrally overlap with the frequency-doubled 1064\,nm science beam. The THG system employs single-pass conversion without a cavity. Although less efficient, this approach is much simpler to implement. Most of the 1596\,nm laser power is directed towards a SHG crystal, where part of the 1596\,nm beam is frequency-doubled to 798\,nm. The output of the SHG would be a mixture of the produced 798\,nm beam and mostly unconverted 1596\,nm beam. They then co-propagate through the SFG crystal, where the 798\,nm photons will combine with the 1596\,nm photons to produce the 532\,nm photons. The output of the SFG crystal will include laser beams at three different wavelengths, with the original 1596\,nm laser dominating the power. 

\begin{figure}[h]
    \centering
    \includegraphics[width=0.6\linewidth]{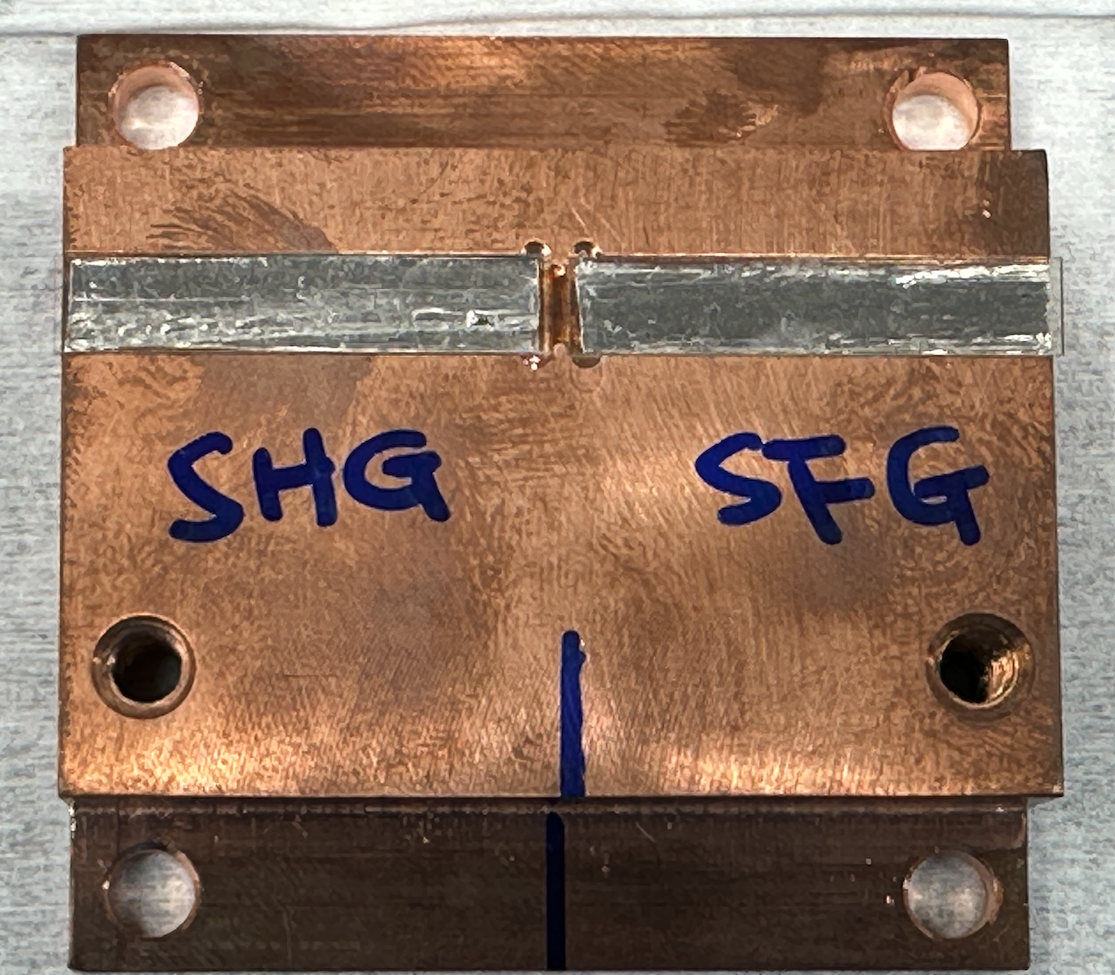}
    \caption{The 1596\,nm THG single-pass copper oven. The SHG and SFG crystals are mounted in series inside the oven, with their alignment determined by the oven machining accuracy. A 1\,mm gap is maintained between the crystals to prevent contact between their AR-coated surfaces. Both crystals are wrapped in Indium foil to enhance thermal coupling and ensure uniform temperature control.}
    \label{fig:oven}
\end{figure}

The 1596\,nm SHG and SFG crystals are acquired from SLF LaserFabriken AB. Both ends of the crystals are anti-reflecting (AR) coated at 1596/798/532\,nm. The crystals have dimensions of 6\,mm × 1\,mm × 30\,mm, and they are designed to have the same phase-matching temperature. This allows them to be mounted in series within a single copper oven and simultaneously controlled at a temperature of 40\,°C. Fig.\ref{fig:oven} shows the oven and crystal assembly, which is mounted on a compact five-axis stage for fine alignment to the input beam. To maximise the THG efficiency, the 1596\,nm beam is focused to a waist of approximately 50\,$\rm \mu m$ at the midpoint between the two crystals to strike a balance between the SHG and SFG processes.  

At the output of the oven, an optical band-pass filter with centre wavelength of 532\,nm (Thorlabs FBW5532-10) filters out the other two co-propagating beams, leaving only the 532\,nm beam. After filtering, we measured about 400\,nW 532\,nm beam with about 1.2\,W original 1596\,nm beam at the input of the oven.

The two 532\,nm beams, one generated from the THG process of the 1596\,nm auxiliary beam and the other from the SHG process of the 1064\,nm science beam, are combined at a beam splitter, where they produce a beat signal. To maximise the beat efficiency, the two beams are carefully aligned and mode-matched to have overlapping waists of approximately 0.5\,mm at the beam splitter. A polarising beam splitter is placed after the combining beam splitter to further suppress any undesired polarisation components. Finally, the beam is received by a 150\,MHz bandwidth silicon amplified photo detector (Thorlabs PDA10A2) to detect the beat signal. 

Although only 400\,nW at 532\,nm was measured via the THG process, a detectable beat signal was obtained due to the sufficient power of the other 532\,nm beam produced via SHG. In practice, 2\,mW (out of 500\,mW available) of SHG power was used to achieve an optimal beat signal when combined with the 400\,nW THG beam. The signal is further amplified and mixed with a frequency-offset oscillator signal (maximum frequency of 160\,MHz). The output of the mixer acts as a phase-locking error signal, and this error signal is low-pass filtered and fed back to the 1596\,nm laser PZT so that the 1596\,nm auxiliary beam is effectively phase-locked to the 1064\,nm science beam, with the frequency offset $\Delta f$ at 532\,nm equal to the frequency-offset signal. 

The other parts of the 1596 nm laser beam and the 1064\,nm science beam are both phase modulated and sent to the science cavity via two dichroic mirrors (Thorlabs DMLP1180) located at both ends of the cavity, which is highly reflective at 1064\,nm and highly transmissive at 1596\,nm. The science cavity is a 400\,$\pm$\,1\,mm long two-mirror linear cavity with the flat mirror glued to a ring PZT for cavity length actuation. The PZT is further glued on a solid aluminium block holder, which has a through hole at the beam height to allow the laser beams to pass. The cavity mirrors are custom-designed and manufactured by Layertec GmbH, and both are dual-wavelength coated with power reflectivity $R=95\,\%$ at 1596\,nm and $R=99\,\%$ at 1064\,nm on the front side and $R<0.2\,\%$ at both wavelengths at the back side. 

Both the auxiliary and science beams are independently monitored via photo detectors at the cavity transmission. The cavity reflected beams are obtained from the rejection ports of the Faraday isolators. Cavity locking error signals can be obtained by using standard PDH locking techniques. An AOM is placed in the 1064\,nm beam path before the science cavity. This setup allows for a frequency offset between the 1064\,nm phase-locking beam and the science cavity locking beam, thereby providing an additional actuator for frequency adjustment.

The ALS system works in the following manner: the cavity will first be locked and stabilised to the 1596\,nm laser with the 1596\,nm PDH control signal fed back to the cavity PZT. By slowly adjusting the phase locking frequency offset between the two 532\,nm beams, the 1064\,nm science beam can be brought to resonate in the cavity. Once the 1064\,nm beam is within the cavity linewidth, the 1064\,nm PDH control signal can be engaged to control the science cavity length PZT while simultaneously the 1596\,nm PDH control signal is disengaged.

\section{Results}
\label{results}
\begin{figure}[h]
    \centering
    \includegraphics[width=0.8\linewidth]{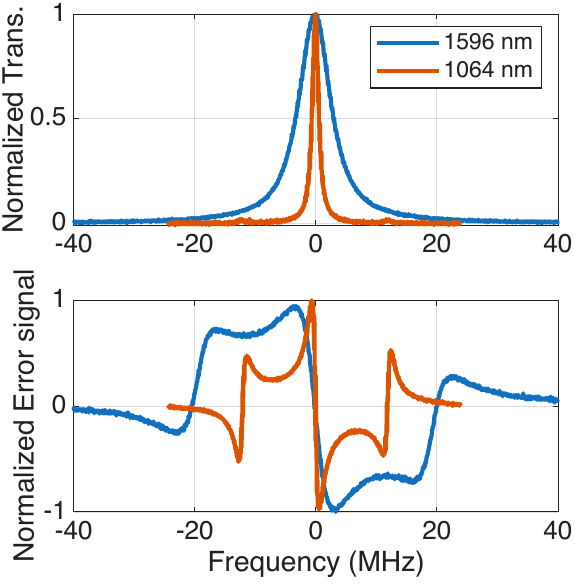}
    \caption{The cavity resonances and PDH error signals with 1596\,nm and 1064\,beams.}
    \label{fig:cavityprofile}
\end{figure}

We first measured the cavity linewidth with both the auxiliary and the science lasers. This was done by scanning the cavity PZT and comparing the cavity linewidth with the modulation sidebands from the PDH error signal. The cavity resonance profiles and the corresponding error signals are shown in Fig.\ref{fig:cavityprofile}. 

The 1596\,nm beam was phase-modulated at 20.2\,MHz, while the 1064\,nm beam was modulated at 12.0\,MHz. The cavity scan was repeated multiple times to estimate the statistical uncertainty. The cavity linewidths were determined to be 6.09\,$\pm$\,0.15\,MHz for the 1596\,nm beam and 1.25\,$\pm$\,0.06\,MHz for the 1064\,nm beam. The cavity FSR is $375\,\pm\,0.9$\,MHz, so the linewidth measurements indicate cavity finesses of 61.6\,$\pm$\,1.6 for the 1596\,nm beam and 299\,$\pm$\,15 for the 1064\,nm beam respectively. These results are in good agreement with the theoretical values predicted from the mirror coating specifications at the corresponding wavelengths.

\begin{figure}[h]
    \centering
    \includegraphics[width=0.9\linewidth]{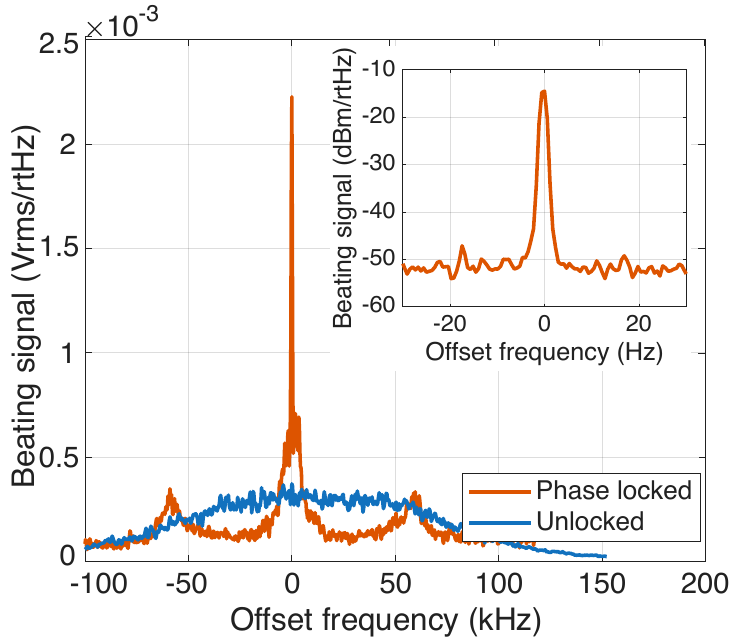}
    \caption{Power spectrum of the 532\,nm beam beat signal around the centre frequency with and without the PLL. The linewidth of the beat signal is around 145\,kHz without the PLL and is reduced to around 1.4\,Hz with the PLL engaged. 
    The inset plot shows a zoomed-in view of the beat note, showing a -3dB linewidth of about 1.4\,Hz.}
    \label{fig:PLL_spctrum}
\end{figure}

The two 532\,nm beams were subsequently phase-locked. They were first combined on a PD with a bandwidth of 150\,MHz ($\rm PD_{532}$ in Fig.\ref{fig:setup}) to detect the beat note between them. Due to imperfections in the SHG locking loops, residual modulation sidebands at 12\,MHz and its harmonics at 24\,MHz leaked into the 532\,nm beat signal. To suppress them, the signal was filtered using two cascaded 50\,MHz high-pass filters (Mini-Circuits BHP-50+). The filtered signal was then demodulated with a 90\,MHz frequency-offset oscillator, serving as the error signal for the PLL. This error signal was applied to 1596\,nm laser PZTs to achieve a stable phase locking between the two 532\,nm beams. The performance of the PLL is shown in Fig.\ref{fig:PLL_spctrum}, with the PLL, the beat signal linewidth was reduced from 145\,kHz to 1.4\,Hz.


\begin{figure}[h]
    \centering
    \includegraphics[width=0.9\linewidth]{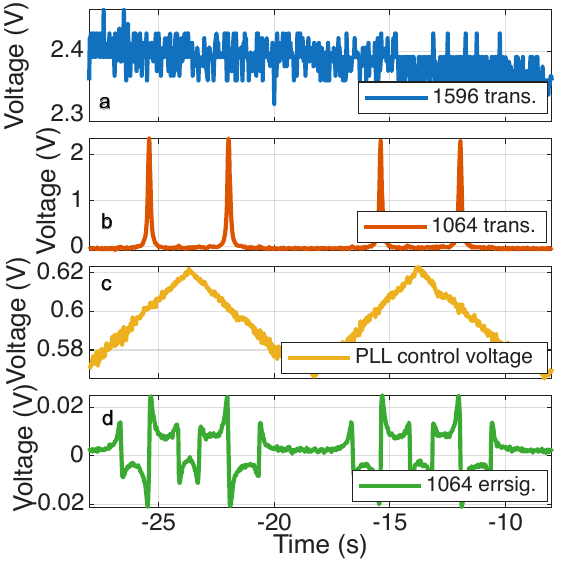}
    \caption{Cavity detuning of the 1064\,nm science laser with the ALS system. Panel (a) - cavity transmission of the 1596\,nm auxiliary beam; panel (b) - cavity transmission of the 1064\,nm science beam; panel (c) - the control voltage that is applied to the 1596\,nm laser PZT through the PLL; panel (d) - the corresponding PDH error signal of the 1064\,nm beam. }
    \label{fig:PLL_scan}
\end{figure}

With the PLL closed, we first demonstrated the capacity of the ALS system to control the detuning of the science laser with respect to the cavity, shown in Fig.\ref{fig:PLL_scan}. In panel (a), the cavity was locked to the 1596\,nm beam, and a steady transmission signal of approximately 2.4\,V was recorded on the 1596\,nm transmission PD ($\rm PD^{1596}_{Trans.}$ in Fig.\ref{fig:setup}). The PLL frequency offset was then scanned with a triangular waveform at 0.1\,Hz , as illustrated in panel (c). The scan of the PLL frequency offset sees the cavity sweep through the resonances of the 1064\,nm science beam, as shown in panel (b), and the corresponding PDH error signal is presented in panel (d). 

In our setup, the maximum output frequency of the signal generator used to produce the frequency offset between the 532\,nm beams, together with the detection bandwidth of the photodetector sensing the 532\,nm beat signal, limited the achievable cavity detuning. In the test shown in Fig.~\ref{fig:PLL_scan}, the cavity was tuned over a range of approximately 45\,MHz around the 1064\,nm resonance by adjusting the PLL offset frequency. For $\rm A^{\#}$ arm cavity with FSR of 37.5\,kHz, the ALS system should be capable of sweeping across multiple FSRs easily.

\begin{figure}[h]
    \centering
    \includegraphics[width=0.9\linewidth]{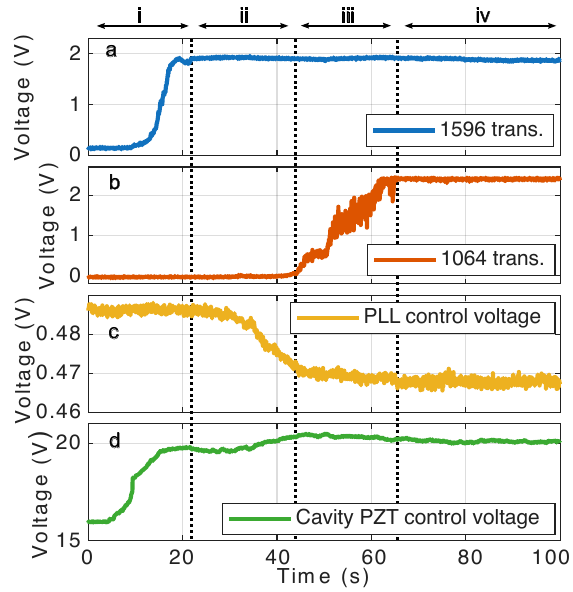}
    \caption{Cavity locking sequence from the 1596\,nm auxiliary laser to the 1064\,nm science laser. Panel (a) - cavity transmission at 1596\,nm; panel (b) - cavity transmission at 1064\,nm; panel (c) - PLL control voltage; panel (d) - the control voltage that is applied to the cavity PZT. The sequence is divided into four stages and is discussed in the main text. The PLL control voltage differs from that shown in Fig.~\ref{fig:PLL_scan} because the two measurement sets were taken at different times.}
    \label{fig:lockingtransion}
\end{figure}

Shown in Fig.\ref{fig:lockingtransion}, we then demonstrated the ALS locking transition from 1596\,nm to 1064\,nm control. The whole demonstrated sequence lasted about 100 seconds and can be divided into four stages:
\begin{itemize}
    \item (i) From 0 to 22\,s, the science cavity was initially set to be off-resonance for both 1596\,nm and 1064 nm beams. Then, the cavity length was manually tuned by actuating the cavity PZT from 15.8\,V to 19.7\,V until the 1596\,nm beam started to resonate inside the cavity, as evidenced by the gradual increase in 1596\,nm transmission power shown in panel (a). During this stage, the PLL control voltage remained stable, and the transmission of the 1064\,nm beam was zero, indicating that it was maintained off-resonance by the PLL. The 1596\,nm cavity locking loop was engaged at 22\,s.
    \item (ii) From 22\,s to 44\,s, while the cavity remained stably locked to the 1596\,nm beam, the PLL frequency offset was adjusted to bring the 1064\,nm beam close to resonance. In response to this adjustment, the PLL control signal decreased from 0.485\,V to 0.47\,V. The cavity PZT control signal increased from 19.7\,V to 20.5\,V accordingly, since the cavity length was changing to accommodate the shift of the change in frequency of the 1596\,nm beam, while maintaining the 1596\,nm transmission.
    \item (iii) From 44\,s to 65\,s, when the cavity was approaching the resonance of the 1064\,nm beam, the PLL frequency offset tuning was slowed down until the maximum resonance resulted in a 1064\,nm transition beam measured voltage of 2.4\,V. At approximately 65\,s, the cavity locking feedback was transferred from the 1596\,nm control loop to the 1064\,nm control loop. This transition was implemented by manually operating the physical switch that selects which feedback signal is applied to the cavity PZT.
    \item (iv) From 65\,s to 100\,s, the cavity is locked exclusively by the 1064\,nm control loop, while both the 1596\,nm and 1064\,nm beams are simultaneously resonant within the cavity. 
\end{itemize}

The above locking transition was successfully repeated multiple times over several days. The lasers were stable enough that it was not necessary to search for the 1064\,nm co-resonance with the 1596\,nm beam each time.  This demonstrates a reliable and robust ALS system.

\section{Discussion and Conclusion}
\label{discuss}
We have demonstrated a novel proof-of-principal ALS system for gravitational wave detectors that use crystalline AlGaAs/GaAs coated mirrors. This new scheme employs an auxiliary laser at 1596\,nm. The 1596\,nm beam is phase locked to the main 1064\,nm beam via the third and second harmonic respectively at 532\,nm. This phase locking between the two 532\,nm beams, reached a 1.4\,Hz linewidth beat signal. If the frequency noise is dominated entirely by the 1596\,nm laser, the frequency noise of the 1596\,nm would be $\delta \nu$ = 0.47\,Hz. If this ALS system is applied to the $\rm A^{\#}$ arm cavities, the frequency noise can be further converted to the cavity length noise $\delta L$ using $\delta \nu/\nu = \delta L/L$, where $\nu$ is the laser frequency and $L$ = 4\,km is the arm cavity length. The estimated length noise is $\sim$10\,pm, which is well below the 1\,nm linewidth of the $\rm A^{\#}$ arm cavity. Furthermore, using a tabletop cavity, we have experimentally demonstrated the stable cavity detuning using the PLL, and stable cavity locking transition from auxiliary laser to the science laser, confirming the viability of this scheme. 

The performance of this novel ALS system could be further improved for long baseline GWDs. 
One limitation is the low conversion efficiency of the single-pass THG process, which results in a relatively weak 532\,nm beam from the auxiliary path. This places stringent requirements on the sensitivity of the photodetector used for beat-note detection and the mode matching and alignment of two beat beams. Future improvements could include the use of a low-noise photodetector to enhance the signal-to-noise ratio of the phase-locking signal. Alternatively, the THG process could be enhanced by incorporating a series of resonant cavities in the same principle as the 1064 nm SHG, which will significantly increase the generated 532\,nm power and relax the detection requirements.

The locking handover is currently achieved via a physical switch that selects the feedback signal applied to the cavity PZT. This abrupt switching may introduce transient excitations and excess control noise. A smoother and more controlled transition, for example, through a continuous blending or gain ramping between the two control loops, would be desirable in future implementations.

Overall, the demonstrated 1064\,nm/1596\,nm ALS scheme provides a viable solution for future gravitational wave detectors employing AlGaAs/GaAs coated mirrors. With further optimisations, the system can be readily extended to more complex interferometric configurations.

\begin{acknowledgments}
This project was funded by the Special Initiative of Australia Research Council Centre of Excellence for Gravitational Wave Discovery (OzGrav CE230100016). S. S. Y. C thanks OzGrav for cross-nodal travel support. The authors thank the LIGO Scientific Collaboration Advanced Interferometer Configuration Working Group for their useful discussions and feedback.
\end{acknowledgments}

\bibliography{apssamp}

\begin{thebibliography}{20}%
\makeatletter
\providecommand \@ifxundefined [1]{%
 \@ifx{#1\undefined}
}%
\providecommand \@ifnum [1]{%
 \ifnum #1\expandafter \@firstoftwo
 \else \expandafter \@secondoftwo
 \fi
}%
\providecommand \@ifx [1]{%
 \ifx #1\expandafter \@firstoftwo
 \else \expandafter \@secondoftwo
 \fi
}%
\providecommand \natexlab [1]{#1}%
\providecommand \enquote  [1]{``#1''}%
\providecommand \bibnamefont  [1]{#1}%
\providecommand \bibfnamefont [1]{#1}%
\providecommand \citenamefont [1]{#1}%
\providecommand \href@noop [0]{\@secondoftwo}%
\providecommand \href [0]{\begingroup \@sanitize@url \@href}%
\providecommand \@href[1]{\@@startlink{#1}\@@href}%
\providecommand \@@href[1]{\endgroup#1\@@endlink}%
\providecommand \@sanitize@url [0]{\catcode `\\12\catcode `\$12\catcode
  `\&12\catcode `\#12\catcode `\^12\catcode `\_12\catcode `\%12\relax}%
\providecommand \@@startlink[1]{}%
\providecommand \@@endlink[0]{}%
\providecommand \url  [0]{\begingroup\@sanitize@url \@url }%
\providecommand \@url [1]{\endgroup\@href {#1}{\urlprefix }}%
\providecommand \urlprefix  [0]{URL }%
\providecommand \Eprint [0]{\href }%
\providecommand \doibase [0]{https://doi.org/}%
\providecommand \selectlanguage [0]{\@gobble}%
\providecommand \bibinfo  [0]{\@secondoftwo}%
\providecommand \bibfield  [0]{\@secondoftwo}%
\providecommand \translation [1]{[#1]}%
\providecommand \BibitemOpen [0]{}%
\providecommand \bibitemStop [0]{}%
\providecommand \bibitemNoStop [0]{.\EOS\space}%
\providecommand \EOS [0]{\spacefactor3000\relax}%
\providecommand \BibitemShut  [1]{\csname bibitem#1\endcsname}%
\let\auto@bib@innerbib\@empty
\bibitem [{\citenamefont {Aasi}\ \emph {et~al.}(2015)\citenamefont {Aasi},
  \citenamefont {Abbott}, \citenamefont {Abbott}, \citenamefont {Abbott},
  \citenamefont {Abernathy}, \citenamefont {Ackley}, \citenamefont {Adams},
  \citenamefont {Adams}, \citenamefont {Addesso}, \citenamefont {Adhikari}
  \emph {et~al.}}]{aasi2015advanced}%
  \BibitemOpen
  \bibfield  {author} {\bibinfo {author} {\bibfnamefont {J.}~\bibnamefont
  {Aasi}}, \bibinfo {author} {\bibfnamefont {B.}~\bibnamefont {Abbott}},
  \bibinfo {author} {\bibfnamefont {R.}~\bibnamefont {Abbott}}, \bibinfo
  {author} {\bibfnamefont {T.}~\bibnamefont {Abbott}}, \bibinfo {author}
  {\bibfnamefont {M.}~\bibnamefont {Abernathy}}, \bibinfo {author}
  {\bibfnamefont {K.}~\bibnamefont {Ackley}}, \bibinfo {author} {\bibfnamefont
  {C.}~\bibnamefont {Adams}}, \bibinfo {author} {\bibfnamefont
  {T.}~\bibnamefont {Adams}}, \bibinfo {author} {\bibfnamefont
  {P.}~\bibnamefont {Addesso}}, \bibinfo {author} {\bibfnamefont
  {R.}~\bibnamefont {Adhikari}}, \emph {et~al.},\ }\bibfield  {title} {\bibinfo
  {title} {Advanced ligo},\ }\href@noop {} {\bibfield  {journal} {\bibinfo
  {journal} {Classical and Quantum Gravity}\ }\textbf {\bibinfo {volume}
  {32}},\ \bibinfo {pages} {074001} (\bibinfo {year} {2015})}\BibitemShut
  {NoStop}%
\bibitem [{\citenamefont {Acernese}\ \emph {et~al.}(2014)\citenamefont
  {Acernese}, \citenamefont {Agathos}, \citenamefont {Agatsuma}, \citenamefont
  {Aisa}, \citenamefont {Allemandou}, \citenamefont {Allocca}, \citenamefont
  {Amarni}, \citenamefont {Astone}, \citenamefont {Balestri}, \citenamefont
  {Ballardin} \emph {et~al.}}]{acernese2014advanced}%
  \BibitemOpen
  \bibfield  {author} {\bibinfo {author} {\bibfnamefont {F.~a.}\ \bibnamefont
  {Acernese}}, \bibinfo {author} {\bibfnamefont {M.}~\bibnamefont {Agathos}},
  \bibinfo {author} {\bibfnamefont {K.}~\bibnamefont {Agatsuma}}, \bibinfo
  {author} {\bibfnamefont {D.}~\bibnamefont {Aisa}}, \bibinfo {author}
  {\bibfnamefont {N.}~\bibnamefont {Allemandou}}, \bibinfo {author}
  {\bibfnamefont {A.}~\bibnamefont {Allocca}}, \bibinfo {author} {\bibfnamefont
  {J.}~\bibnamefont {Amarni}}, \bibinfo {author} {\bibfnamefont
  {P.}~\bibnamefont {Astone}}, \bibinfo {author} {\bibfnamefont
  {G.}~\bibnamefont {Balestri}}, \bibinfo {author} {\bibfnamefont
  {G.}~\bibnamefont {Ballardin}}, \emph {et~al.},\ }\bibfield  {title}
  {\bibinfo {title} {Advanced virgo: a second-generation interferometric
  gravitational wave detector},\ }\href@noop {} {\bibfield  {journal} {\bibinfo
   {journal} {Classical and Quantum Gravity}\ }\textbf {\bibinfo {volume}
  {32}},\ \bibinfo {pages} {024001} (\bibinfo {year} {2014})}\BibitemShut
  {NoStop}%
\bibitem [{\citenamefont {Aso}\ \emph {et~al.}(2013)\citenamefont {Aso},
  \citenamefont {Michimura}, \citenamefont {Somiya}, \citenamefont {Ando},
  \citenamefont {Miyakawa}, \citenamefont {Sekiguchi}, \citenamefont {Tatsumi},
  \citenamefont {Yamamoto}, \citenamefont {Collaboration} \emph
  {et~al.}}]{aso2013interferometer}%
  \BibitemOpen
  \bibfield  {author} {\bibinfo {author} {\bibfnamefont {Y.}~\bibnamefont
  {Aso}}, \bibinfo {author} {\bibfnamefont {Y.}~\bibnamefont {Michimura}},
  \bibinfo {author} {\bibfnamefont {K.}~\bibnamefont {Somiya}}, \bibinfo
  {author} {\bibfnamefont {M.}~\bibnamefont {Ando}}, \bibinfo {author}
  {\bibfnamefont {O.}~\bibnamefont {Miyakawa}}, \bibinfo {author}
  {\bibfnamefont {T.}~\bibnamefont {Sekiguchi}}, \bibinfo {author}
  {\bibfnamefont {D.}~\bibnamefont {Tatsumi}}, \bibinfo {author} {\bibfnamefont
  {H.}~\bibnamefont {Yamamoto}}, \bibinfo {author} {\bibfnamefont
  {K.}~\bibnamefont {Collaboration}}, \emph {et~al.},\ }\bibfield  {title}
  {\bibinfo {title} {Interferometer design of the kagra gravitational wave
  detector},\ }\href@noop {} {\bibfield  {journal} {\bibinfo  {journal}
  {Physical Review D}\ }\textbf {\bibinfo {volume} {88}},\ \bibinfo {pages}
  {043007} (\bibinfo {year} {2013})}\BibitemShut {NoStop}%
\bibitem [{\citenamefont {Staley}\ \emph {et~al.}(2014)\citenamefont {Staley},
  \citenamefont {Martynov}, \citenamefont {Abbott}, \citenamefont {Adhikari},
  \citenamefont {Arai}, \citenamefont {Ballmer}, \citenamefont {Barsotti},
  \citenamefont {Brooks}, \citenamefont {DeRosa}, \citenamefont {Dwyer} \emph
  {et~al.}}]{staley2014achieving}%
  \BibitemOpen
  \bibfield  {author} {\bibinfo {author} {\bibfnamefont {A.}~\bibnamefont
  {Staley}}, \bibinfo {author} {\bibfnamefont {D.}~\bibnamefont {Martynov}},
  \bibinfo {author} {\bibfnamefont {R.}~\bibnamefont {Abbott}}, \bibinfo
  {author} {\bibfnamefont {R.}~\bibnamefont {Adhikari}}, \bibinfo {author}
  {\bibfnamefont {K.}~\bibnamefont {Arai}}, \bibinfo {author} {\bibfnamefont
  {S.}~\bibnamefont {Ballmer}}, \bibinfo {author} {\bibfnamefont
  {L.}~\bibnamefont {Barsotti}}, \bibinfo {author} {\bibfnamefont
  {A.}~\bibnamefont {Brooks}}, \bibinfo {author} {\bibfnamefont
  {R.}~\bibnamefont {DeRosa}}, \bibinfo {author} {\bibfnamefont
  {S.}~\bibnamefont {Dwyer}}, \emph {et~al.},\ }\bibfield  {title} {\bibinfo
  {title} {Achieving resonance in the advanced ligo gravitational-wave
  interferometer},\ }\href@noop {} {\bibfield  {journal} {\bibinfo  {journal}
  {Classical and Quantum Gravity}\ }\textbf {\bibinfo {volume} {31}},\ \bibinfo
  {pages} {245010} (\bibinfo {year} {2014})}\BibitemShut {NoStop}%
\bibitem [{\citenamefont {Mullavey}\ \emph {et~al.}(2011)\citenamefont
  {Mullavey}, \citenamefont {Slagmolen}, \citenamefont {Miller}, \citenamefont
  {Evans}, \citenamefont {Fritschel}, \citenamefont {Sigg}, \citenamefont
  {Waldman}, \citenamefont {Shaddock},\ and\ \citenamefont
  {McClelland}}]{mullavey2011arm}%
  \BibitemOpen
  \bibfield  {author} {\bibinfo {author} {\bibfnamefont {A.~J.}\ \bibnamefont
  {Mullavey}}, \bibinfo {author} {\bibfnamefont {B.~J.}\ \bibnamefont
  {Slagmolen}}, \bibinfo {author} {\bibfnamefont {J.}~\bibnamefont {Miller}},
  \bibinfo {author} {\bibfnamefont {M.}~\bibnamefont {Evans}}, \bibinfo
  {author} {\bibfnamefont {P.}~\bibnamefont {Fritschel}}, \bibinfo {author}
  {\bibfnamefont {D.}~\bibnamefont {Sigg}}, \bibinfo {author} {\bibfnamefont
  {S.~J.}\ \bibnamefont {Waldman}}, \bibinfo {author} {\bibfnamefont {D.~A.}\
  \bibnamefont {Shaddock}},\ and\ \bibinfo {author} {\bibfnamefont {D.~E.}\
  \bibnamefont {McClelland}},\ }\bibfield  {title} {\bibinfo {title}
  {Arm-length stabilisation for interferometric gravitational-wave detectors
  using frequency-doubled auxiliary lasers},\ }\href@noop {} {\bibfield
  {journal} {\bibinfo  {journal} {Optics Express}\ }\textbf {\bibinfo {volume}
  {20}},\ \bibinfo {pages} {81} (\bibinfo {year} {2011})}\BibitemShut {NoStop}%
\bibitem [{\citenamefont {Akutsu}\ \emph {et~al.}(2020)\citenamefont {Akutsu},
  \citenamefont {Ando}, \citenamefont {Arai}, \citenamefont {Arai},
  \citenamefont {Arai}, \citenamefont {Araki}, \citenamefont {Araya},
  \citenamefont {Aritomi}, \citenamefont {Aso}, \citenamefont {Bae} \emph
  {et~al.}}]{akutsu2020arm}%
  \BibitemOpen
  \bibfield  {author} {\bibinfo {author} {\bibfnamefont {T.}~\bibnamefont
  {Akutsu}}, \bibinfo {author} {\bibfnamefont {M.}~\bibnamefont {Ando}},
  \bibinfo {author} {\bibfnamefont {K.}~\bibnamefont {Arai}}, \bibinfo {author}
  {\bibfnamefont {K.}~\bibnamefont {Arai}}, \bibinfo {author} {\bibfnamefont
  {Y.}~\bibnamefont {Arai}}, \bibinfo {author} {\bibfnamefont {S.}~\bibnamefont
  {Araki}}, \bibinfo {author} {\bibfnamefont {A.}~\bibnamefont {Araya}},
  \bibinfo {author} {\bibfnamefont {N.}~\bibnamefont {Aritomi}}, \bibinfo
  {author} {\bibfnamefont {Y.}~\bibnamefont {Aso}}, \bibinfo {author}
  {\bibfnamefont {S.}~\bibnamefont {Bae}}, \emph {et~al.},\ }\bibfield  {title}
  {\bibinfo {title} {An arm length stabilization system for kagra and future
  gravitational-wave detectors},\ }\href@noop {} {\bibfield  {journal}
  {\bibinfo  {journal} {Classical and Quantum Gravity}\ }\textbf {\bibinfo
  {volume} {37}},\ \bibinfo {pages} {035004} (\bibinfo {year}
  {2020})}\BibitemShut {NoStop}%
\bibitem [{\citenamefont {Izumi}\ \emph {et~al.}(2012)\citenamefont {Izumi},
  \citenamefont {Arai}, \citenamefont {Barr}, \citenamefont {Betzwieser},
  \citenamefont {Brooks}, \citenamefont {Dahl}, \citenamefont {Doravari},
  \citenamefont {Driggers}, \citenamefont {Korth}, \citenamefont {Miao} \emph
  {et~al.}}]{izumi2012multicolor}%
  \BibitemOpen
  \bibfield  {author} {\bibinfo {author} {\bibfnamefont {K.}~\bibnamefont
  {Izumi}}, \bibinfo {author} {\bibfnamefont {K.}~\bibnamefont {Arai}},
  \bibinfo {author} {\bibfnamefont {B.}~\bibnamefont {Barr}}, \bibinfo {author}
  {\bibfnamefont {J.}~\bibnamefont {Betzwieser}}, \bibinfo {author}
  {\bibfnamefont {A.}~\bibnamefont {Brooks}}, \bibinfo {author} {\bibfnamefont
  {K.}~\bibnamefont {Dahl}}, \bibinfo {author} {\bibfnamefont {S.}~\bibnamefont
  {Doravari}}, \bibinfo {author} {\bibfnamefont {J.~C.}\ \bibnamefont
  {Driggers}}, \bibinfo {author} {\bibfnamefont {W.~Z.}\ \bibnamefont {Korth}},
  \bibinfo {author} {\bibfnamefont {H.}~\bibnamefont {Miao}}, \emph {et~al.},\
  }\bibfield  {title} {\bibinfo {title} {Multicolor cavity metrology},\
  }\href@noop {} {\bibfield  {journal} {\bibinfo  {journal} {Journal of the
  Optical Society of America A}\ }\textbf {\bibinfo {volume} {29}},\ \bibinfo
  {pages} {2092} (\bibinfo {year} {2012})}\BibitemShut {NoStop}%
\bibitem [{\citenamefont {Drever}\ \emph {et~al.}(1983)\citenamefont {Drever},
  \citenamefont {Hall}, \citenamefont {Kowalski}, \citenamefont {Hough},
  \citenamefont {Ford}, \citenamefont {Munley},\ and\ \citenamefont
  {Ward}}]{drever1983laser}%
  \BibitemOpen
  \bibfield  {author} {\bibinfo {author} {\bibfnamefont {R.~W.}\ \bibnamefont
  {Drever}}, \bibinfo {author} {\bibfnamefont {J.~L.}\ \bibnamefont {Hall}},
  \bibinfo {author} {\bibfnamefont {F.~V.}\ \bibnamefont {Kowalski}}, \bibinfo
  {author} {\bibfnamefont {J.}~\bibnamefont {Hough}}, \bibinfo {author}
  {\bibfnamefont {G.}~\bibnamefont {Ford}}, \bibinfo {author} {\bibfnamefont
  {A.}~\bibnamefont {Munley}},\ and\ \bibinfo {author} {\bibfnamefont
  {H.}~\bibnamefont {Ward}},\ }\bibfield  {title} {\bibinfo {title} {Laser
  phase and frequency stabilization using an optical resonator},\ }\href@noop
  {} {\bibfield  {journal} {\bibinfo  {journal} {Applied Physics B}\ }\textbf
  {\bibinfo {volume} {31}},\ \bibinfo {pages} {97} (\bibinfo {year}
  {1983})}\BibitemShut {NoStop}%
\bibitem [{\citenamefont {Capote}\ \emph {et~al.}(2025)\citenamefont {Capote},
  \citenamefont {Jia}, \citenamefont {Aritomi}, \citenamefont {Nakano},
  \citenamefont {Xu}, \citenamefont {Abbott}, \citenamefont {Abouelfettouh},
  \citenamefont {Adhikari}, \citenamefont {Ananyeva}, \citenamefont {Appert}
  \emph {et~al.}}]{capote2025advanced}%
  \BibitemOpen
  \bibfield  {author} {\bibinfo {author} {\bibfnamefont {E.}~\bibnamefont
  {Capote}}, \bibinfo {author} {\bibfnamefont {W.}~\bibnamefont {Jia}},
  \bibinfo {author} {\bibfnamefont {N.}~\bibnamefont {Aritomi}}, \bibinfo
  {author} {\bibfnamefont {M.}~\bibnamefont {Nakano}}, \bibinfo {author}
  {\bibfnamefont {V.}~\bibnamefont {Xu}}, \bibinfo {author} {\bibfnamefont
  {R.}~\bibnamefont {Abbott}}, \bibinfo {author} {\bibfnamefont
  {I.}~\bibnamefont {Abouelfettouh}}, \bibinfo {author} {\bibfnamefont
  {R.}~\bibnamefont {Adhikari}}, \bibinfo {author} {\bibfnamefont
  {A.}~\bibnamefont {Ananyeva}}, \bibinfo {author} {\bibfnamefont
  {S.}~\bibnamefont {Appert}}, \emph {et~al.},\ }\bibfield  {title} {\bibinfo
  {title} {Advanced ligo detector performance in the fourth observing run},\
  }\href@noop {} {\bibfield  {journal} {\bibinfo  {journal} {Physical Review
  D}\ }\textbf {\bibinfo {volume} {111}},\ \bibinfo {pages} {062002} (\bibinfo
  {year} {2025})}\BibitemShut {NoStop}%
\bibitem [{\citenamefont {Cole}\ \emph {et~al.}(2013)\citenamefont {Cole},
  \citenamefont {Zhang}, \citenamefont {Martin}, \citenamefont {Ye},\ and\
  \citenamefont {Aspelmeyer}}]{cole2013tenfold}%
  \BibitemOpen
  \bibfield  {author} {\bibinfo {author} {\bibfnamefont {G.~D.}\ \bibnamefont
  {Cole}}, \bibinfo {author} {\bibfnamefont {W.}~\bibnamefont {Zhang}},
  \bibinfo {author} {\bibfnamefont {M.~J.}\ \bibnamefont {Martin}}, \bibinfo
  {author} {\bibfnamefont {J.}~\bibnamefont {Ye}},\ and\ \bibinfo {author}
  {\bibfnamefont {M.}~\bibnamefont {Aspelmeyer}},\ }\bibfield  {title}
  {\bibinfo {title} {Tenfold reduction of brownian noise in high-reflectivity
  optical coatings},\ }\href@noop {} {\bibfield  {journal} {\bibinfo  {journal}
  {Nature Photonics}\ }\textbf {\bibinfo {volume} {7}},\ \bibinfo {pages} {644}
  (\bibinfo {year} {2013})}\BibitemShut {NoStop}%
\bibitem [{\citenamefont {Cole}\ \emph {et~al.}(2023)\citenamefont {Cole},
  \citenamefont {Ballmer}, \citenamefont {Billingsley}, \citenamefont
  {Cata{\~n}o-Lopez}, \citenamefont {Fejer}, \citenamefont {Fritschel},
  \citenamefont {Gretarsson}, \citenamefont {Harry}, \citenamefont {Kedar},
  \citenamefont {Legero} \emph {et~al.}}]{cole2023substrate}%
  \BibitemOpen
  \bibfield  {author} {\bibinfo {author} {\bibfnamefont {G.~D.}\ \bibnamefont
  {Cole}}, \bibinfo {author} {\bibfnamefont {S.}~\bibnamefont {Ballmer}},
  \bibinfo {author} {\bibfnamefont {G.}~\bibnamefont {Billingsley}}, \bibinfo
  {author} {\bibfnamefont {S.}~\bibnamefont {Cata{\~n}o-Lopez}}, \bibinfo
  {author} {\bibfnamefont {M.}~\bibnamefont {Fejer}}, \bibinfo {author}
  {\bibfnamefont {P.}~\bibnamefont {Fritschel}}, \bibinfo {author}
  {\bibfnamefont {A.}~\bibnamefont {Gretarsson}}, \bibinfo {author}
  {\bibfnamefont {G.}~\bibnamefont {Harry}}, \bibinfo {author} {\bibfnamefont
  {D.}~\bibnamefont {Kedar}}, \bibinfo {author} {\bibfnamefont
  {T.}~\bibnamefont {Legero}}, \emph {et~al.},\ }\bibfield  {title} {\bibinfo
  {title} {Substrate-transferred gaas/algaas crystalline coatings for
  gravitational-wave detectors},\ }\href@noop {} {\bibfield  {journal}
  {\bibinfo  {journal} {Applied Physics Letters}\ }\textbf {\bibinfo {volume}
  {122}} (\bibinfo {year} {2023})}\BibitemShut {NoStop}%
\bibitem [{\citenamefont {Harry}\ \emph {et~al.}(2007)\citenamefont {Harry},
  \citenamefont {Abernathy}, \citenamefont {Becerra-Toledo}, \citenamefont
  {Armandula}, \citenamefont {Black}, \citenamefont {Dooley}, \citenamefont
  {Eichenfield}, \citenamefont {Nwabugwu}, \citenamefont {Villar},
  \citenamefont {Crooks} \emph {et~al.}}]{harry2007titania}%
  \BibitemOpen
  \bibfield  {author} {\bibinfo {author} {\bibfnamefont {G.~M.}\ \bibnamefont
  {Harry}}, \bibinfo {author} {\bibfnamefont {M.~R.}\ \bibnamefont
  {Abernathy}}, \bibinfo {author} {\bibfnamefont {A.~E.}\ \bibnamefont
  {Becerra-Toledo}}, \bibinfo {author} {\bibfnamefont {H.}~\bibnamefont
  {Armandula}}, \bibinfo {author} {\bibfnamefont {E.}~\bibnamefont {Black}},
  \bibinfo {author} {\bibfnamefont {K.}~\bibnamefont {Dooley}}, \bibinfo
  {author} {\bibfnamefont {M.}~\bibnamefont {Eichenfield}}, \bibinfo {author}
  {\bibfnamefont {C.}~\bibnamefont {Nwabugwu}}, \bibinfo {author}
  {\bibfnamefont {A.}~\bibnamefont {Villar}}, \bibinfo {author} {\bibfnamefont
  {D.}~\bibnamefont {Crooks}}, \emph {et~al.},\ }\bibfield  {title} {\bibinfo
  {title} {Titania-doped tantala/silica coatings for gravitational-wave
  detection},\ }\href@noop {} {\bibfield  {journal} {\bibinfo  {journal}
  {Classical and Quantum Gravity}\ }\textbf {\bibinfo {volume} {24}},\ \bibinfo
  {pages} {405} (\bibinfo {year} {2007})}\BibitemShut {NoStop}%
\bibitem [{\citenamefont {Gupta}\ \emph {et~al.}(2024)\citenamefont {Gupta},
  \citenamefont {Afle}, \citenamefont {Arun}, \citenamefont {Bandopadhyay},
  \citenamefont {Baryakhtar}, \citenamefont {Biscoveanu}, \citenamefont
  {Borhanian}, \citenamefont {Broekgaarden}, \citenamefont {Corsi},
  \citenamefont {Dhani} \emph {et~al.}}]{gupta2024characterizing}%
  \BibitemOpen
  \bibfield  {author} {\bibinfo {author} {\bibfnamefont {I.}~\bibnamefont
  {Gupta}}, \bibinfo {author} {\bibfnamefont {C.}~\bibnamefont {Afle}},
  \bibinfo {author} {\bibfnamefont {K.}~\bibnamefont {Arun}}, \bibinfo {author}
  {\bibfnamefont {A.}~\bibnamefont {Bandopadhyay}}, \bibinfo {author}
  {\bibfnamefont {M.}~\bibnamefont {Baryakhtar}}, \bibinfo {author}
  {\bibfnamefont {S.}~\bibnamefont {Biscoveanu}}, \bibinfo {author}
  {\bibfnamefont {S.}~\bibnamefont {Borhanian}}, \bibinfo {author}
  {\bibfnamefont {F.}~\bibnamefont {Broekgaarden}}, \bibinfo {author}
  {\bibfnamefont {A.}~\bibnamefont {Corsi}}, \bibinfo {author} {\bibfnamefont
  {A.}~\bibnamefont {Dhani}}, \emph {et~al.},\ }\bibfield  {title} {\bibinfo
  {title} {Characterizing gravitational wave detector networks: from A\# to
  cosmic explorer},\ }\href@noop {} {\bibfield  {journal} {\bibinfo  {journal}
  {Classical and Quantum Gravity}\ }\textbf {\bibinfo {volume} {41}},\ \bibinfo
  {pages} {245001} (\bibinfo {year} {2024})}\BibitemShut {NoStop}%
\bibitem [{\citenamefont {Evans}\ \emph {et~al.}(2021)\citenamefont {Evans},
  \citenamefont {Adhikari}, \citenamefont {Afle}, \citenamefont {Ballmer},
  \citenamefont {Biscoveanu}, \citenamefont {Borhanian}, \citenamefont {Brown},
  \citenamefont {Chen}, \citenamefont {Eisenstein}, \citenamefont {Gruson}
  \emph {et~al.}}]{evans2021horizon}%
  \BibitemOpen
  \bibfield  {author} {\bibinfo {author} {\bibfnamefont {M.}~\bibnamefont
  {Evans}}, \bibinfo {author} {\bibfnamefont {R.~X.}\ \bibnamefont {Adhikari}},
  \bibinfo {author} {\bibfnamefont {C.}~\bibnamefont {Afle}}, \bibinfo {author}
  {\bibfnamefont {S.~W.}\ \bibnamefont {Ballmer}}, \bibinfo {author}
  {\bibfnamefont {S.}~\bibnamefont {Biscoveanu}}, \bibinfo {author}
  {\bibfnamefont {S.}~\bibnamefont {Borhanian}}, \bibinfo {author}
  {\bibfnamefont {D.~A.}\ \bibnamefont {Brown}}, \bibinfo {author}
  {\bibfnamefont {Y.}~\bibnamefont {Chen}}, \bibinfo {author} {\bibfnamefont
  {R.}~\bibnamefont {Eisenstein}}, \bibinfo {author} {\bibfnamefont
  {A.}~\bibnamefont {Gruson}}, \emph {et~al.},\ }\bibfield  {title} {\bibinfo
  {title} {A horizon study for cosmic explorer: science, observatories, and
  community},\ }\href@noop {} {\bibfield  {journal} {\bibinfo  {journal} {arXiv
  preprint arXiv:2109.09882}\ } (\bibinfo {year} {2021})}\BibitemShut {NoStop}%
\bibitem [{\citenamefont {Punturo}\ \emph {et~al.}(2010)\citenamefont
  {Punturo}, \citenamefont {Abernathy}, \citenamefont {Acernese}, \citenamefont
  {Allen}, \citenamefont {Andersson}, \citenamefont {Arun}, \citenamefont
  {Barone}, \citenamefont {Barr}, \citenamefont {Barsuglia}, \citenamefont
  {Beker} \emph {et~al.}}]{punturo2010einstein}%
  \BibitemOpen
  \bibfield  {author} {\bibinfo {author} {\bibfnamefont {M.}~\bibnamefont
  {Punturo}}, \bibinfo {author} {\bibfnamefont {M.}~\bibnamefont {Abernathy}},
  \bibinfo {author} {\bibfnamefont {F.}~\bibnamefont {Acernese}}, \bibinfo
  {author} {\bibfnamefont {B.}~\bibnamefont {Allen}}, \bibinfo {author}
  {\bibfnamefont {N.}~\bibnamefont {Andersson}}, \bibinfo {author}
  {\bibfnamefont {K.}~\bibnamefont {Arun}}, \bibinfo {author} {\bibfnamefont
  {F.}~\bibnamefont {Barone}}, \bibinfo {author} {\bibfnamefont
  {B.}~\bibnamefont {Barr}}, \bibinfo {author} {\bibfnamefont {M.}~\bibnamefont
  {Barsuglia}}, \bibinfo {author} {\bibfnamefont {M.}~\bibnamefont {Beker}},
  \emph {et~al.},\ }\bibfield  {title} {\bibinfo {title} {The einstein
  telescope: A third-generation gravitational wave observatory},\ }\href@noop
  {} {\bibfield  {journal} {\bibinfo  {journal} {Classical and Quantum
  Gravity}\ }\textbf {\bibinfo {volume} {27}},\ \bibinfo {pages} {194002}
  (\bibinfo {year} {2010})}\BibitemShut {NoStop}%
\bibitem [{\citenamefont {Blakemore}(1982)}]{blakemore1982semiconducting}%
  \BibitemOpen
  \bibfield  {author} {\bibinfo {author} {\bibfnamefont {J.}~\bibnamefont
  {Blakemore}},\ }\bibfield  {title} {\bibinfo {title} {Semiconducting and
  other major properties of gallium arsenide},\ }\href@noop {} {\bibfield
  {journal} {\bibinfo  {journal} {Journal of Applied Physics}\ }\textbf
  {\bibinfo {volume} {53}},\ \bibinfo {pages} {R123} (\bibinfo {year}
  {1982})}\BibitemShut {NoStop}%
\bibitem [{\citenamefont {Tanioka}\ \emph {et~al.}(2024)\citenamefont
  {Tanioka}, \citenamefont {Wu},\ and\ \citenamefont
  {Ballmer}}]{tanioka2024experimental}%
  \BibitemOpen
  \bibfield  {author} {\bibinfo {author} {\bibfnamefont {S.}~\bibnamefont
  {Tanioka}}, \bibinfo {author} {\bibfnamefont {B.}~\bibnamefont {Wu}},\ and\
  \bibinfo {author} {\bibfnamefont {S.~W.}\ \bibnamefont {Ballmer}},\
  }\bibfield  {title} {\bibinfo {title} {Experimental demonstration of
  frequency-downconverted arm-length stabilization for a future upgraded
  gravitational wave detector},\ }\href@noop {} {\bibfield  {journal} {\bibinfo
   {journal} {Optics Letters}\ }\textbf {\bibinfo {volume} {49}},\ \bibinfo
  {pages} {5763} (\bibinfo {year} {2024})}\BibitemShut {NoStop}%
\bibitem [{\citenamefont {Wang}\ \emph {et~al.}(2023)\citenamefont {Wang},
  \citenamefont {Ma}, \citenamefont {Mei}, \citenamefont {Ji}, \citenamefont
  {Su}, \citenamefont {Zhao}, \citenamefont {Xiao},\ and\ \citenamefont
  {Jia}}]{wang2023simple}%
  \BibitemOpen
  \bibfield  {author} {\bibinfo {author} {\bibfnamefont {F.}~\bibnamefont
  {Wang}}, \bibinfo {author} {\bibfnamefont {W.}~\bibnamefont {Ma}}, \bibinfo
  {author} {\bibfnamefont {F.}~\bibnamefont {Mei}}, \bibinfo {author}
  {\bibfnamefont {Z.}~\bibnamefont {Ji}}, \bibinfo {author} {\bibfnamefont
  {D.}~\bibnamefont {Su}}, \bibinfo {author} {\bibfnamefont {Y.}~\bibnamefont
  {Zhao}}, \bibinfo {author} {\bibfnamefont {L.}~\bibnamefont {Xiao}},\ and\
  \bibinfo {author} {\bibfnamefont {S.}~\bibnamefont {Jia}},\ }\bibfield
  {title} {\bibinfo {title} {Simple, low-cost, and well-performing optical
  phase-locked loop for frequency and phase locking of semiconductor lasers},\
  }\href@noop {} {\bibfield  {journal} {\bibinfo  {journal} {Applied Optics}\
  }\textbf {\bibinfo {volume} {62}},\ \bibinfo {pages} {7169} (\bibinfo {year}
  {2023})}\BibitemShut {NoStop}%
\bibitem [{\citenamefont {Kulur~Ramamohan}\ \emph {et~al.}(2024)\citenamefont
  {Kulur~Ramamohan}, \citenamefont {Chua}, \citenamefont {Zhang}, \citenamefont
  {Yap}, \citenamefont {Wright}, \citenamefont {Holland}, \citenamefont
  {Forsyth},\ and\ \citenamefont {Slagmolen}}]{kulur2024characterization}%
  \BibitemOpen
  \bibfield  {author} {\bibinfo {author} {\bibfnamefont {A.}~\bibnamefont
  {Kulur~Ramamohan}}, \bibinfo {author} {\bibfnamefont {S.}~\bibnamefont
  {Chua}}, \bibinfo {author} {\bibfnamefont {Y.}~\bibnamefont {Zhang}},
  \bibinfo {author} {\bibfnamefont {M.}~\bibnamefont {Yap}}, \bibinfo {author}
  {\bibfnamefont {J.}~\bibnamefont {Wright}}, \bibinfo {author} {\bibfnamefont
  {N.}~\bibnamefont {Holland}}, \bibinfo {author} {\bibfnamefont
  {P.}~\bibnamefont {Forsyth}},\ and\ \bibinfo {author} {\bibfnamefont
  {B.}~\bibnamefont {Slagmolen}},\ }\bibfield  {title} {\bibinfo {title}
  {Characterization of heterodyne optical phase locking for relative laser
  frequency noise suppression in differential measurement},\ }\href@noop {}
  {\bibfield  {journal} {\bibinfo  {journal} {Optics Express}\ }\textbf
  {\bibinfo {volume} {32}},\ \bibinfo {pages} {39793} (\bibinfo {year}
  {2024})}\BibitemShut {NoStop}%
\bibitem [{\citenamefont {Yeaton-Massey}\ and\ \citenamefont
  {Adhikari}(2012)}]{yeaton2012new}%
  \BibitemOpen
  \bibfield  {author} {\bibinfo {author} {\bibfnamefont {D.}~\bibnamefont
  {Yeaton-Massey}}\ and\ \bibinfo {author} {\bibfnamefont {R.~X.}\ \bibnamefont
  {Adhikari}},\ }\bibfield  {title} {\bibinfo {title} {A new bound on excess
  frequency noise in second harmonic generation in ppktp at the 10- 19 level},\
  }\href@noop {} {\bibfield  {journal} {\bibinfo  {journal} {Optics Express}\
  }\textbf {\bibinfo {volume} {20}},\ \bibinfo {pages} {21019} (\bibinfo {year}
  {2012})}\BibitemShut {NoStop}%
\end{thebibliography}%

\end{document}